\documentstyle[aps,prd,psfig]{revtex}

\def\be{\begin{equation}}
\def\ee{\end{equation}}
\def\bea{\begin{eqnarray}}
\def\eea{\end{eqnarray}}

\begin{document}
\draft

\title{Removing non-stationary, non-harmonic external
  interference from 
gravitational wave interferometer data}

\author{Alicia M. Sintes and  Bernard F. Schutz}
\address{Max-Planck-Institut f\"ur Gravitationsphysik
(Albert-Einstein-Institut) \\
Schlaatzweg 1, 14473 Potsdam, Germany}
\date{\today}
 \maketitle

\begin{abstract} 
We describe a procedure  to identify and remove 
a class of non-stationary and non-harmonic interference lines
 from 
gravitational wave interferometer data. These lines appear to be
associated with  the 
external  electricity main supply, but  their amplitudes are 
non-stationary and they do not appear at harmonics of the fundamental
supply frequency.
We find  an empirical model  able to represent coherently
 all the non-harmonic
lines we have found  in the power spectrum, in terms of an assumed
reference signal of the \lq primary' supply input
signal.
If this signal is  not available then 
 it can be reconstructed
from the same data by making use of the coherent line removal algorithm
that we have described elsewhere.
All these lines are broadened by  frequency 
changes  of the supply signal, and they corrupt significant
 frequency ranges of
the power spectrum.
The physical process that generates this interference is so far
unknown, but it is highly non-linear and non-stationary.
Using our model, we cancel the interference in the time domain by 
 an adaptive procedure that should work regardless of the source of
 the primary interference.
 We have applied the method to laser interferometer data from the
 Glasgow prototype detector,
  where all the features
we  describe in this paper were observed.
The algorithm has been tuned in such a way that the entire series of
wide lines corresponding to the electrical interference are removed,
leaving the spectrum clean enough  to detect 
signals previously 
 masked
by them.  Single-line  signals buried in the interference can be
recovered with at least  $75\, \%$ of their original signal amplitude.
 
\end{abstract} 

\section{Introduction} 
Gravitational wave ({\sc gw}) research started in the early 1960's 
thanks to the pioneering work of Weber \cite{W}. Since that time, there
has been an ongoing research effort to develop detectors of sufficient
sensitivity to allow the detection of these waves from astrophysical sources.

The effect of a {\sc gw} of amplitude $h$ is to produce a strain in 
space  given by $\Delta L/L=h/2$. The magnitude of the problem facing
researchers in this area can be appreciate if from the fact that theory
predicts
that for a reasonable 
\lq event rate' one should aim for a strain sensitivity of $10^{-21}$
to  $10^{-22}$. This means  that if we were monitoring the separation
of two free test masses of one meter apart, the change in their separation
would be $10^{-21}$ m. Such figures show the size
 of the experimental challenge facing those developing
gravitational wave detectors, and make it clear that the analysis of the
data must realize as much of the detector sensitivity as possible.

The different types of detectors can be classified into two major
categories: those using laser interferometers with very long arms
\cite{Sa} and those using resonant solid masses that may be cooled to
ultra-low temperatures \cite{Gibbons71,Astone93}. The first  have the ability
to measure the gravitational wave induced strain in a broad frequency band
(expected to range from 50 Hz up to perhaps 5 kHz), while the latter measure
the gravitational wave Fourier components around the  resonant 
frequency (usually near 1 kHz), with a bandwidth currently of order a few Hz.
For resonant mass antennas, the fundamental limitation to their sensitivity
comes from the thermal motion of the atoms that can be reduced by cooling
them to temperatures of order 50 mK; existing antennas can then achieve a 
sensitivity  $10^{-19}$
to  $10^{-20}$. 

In the early 1970's, the idea emerged that laser interferometers might have
a better chance of detecting gravitational waves. Detailed studies were
carried out by Forward and his group \cite{Fo} and by  Weiss \cite{Wei}.
Since then, several groups have develop prototype interferometric detectors
at Glasgow (10 meter Fabry-Perot), Garching (30 m delay line), MIT,
Caltech (40 m Fabry-Perot) and Tokyo. 
Projects to build long arm laser interferometers have also been funded.
These are  LIGO project \cite{Abra1} to build two 4 km detectors in USA, 
VIRGO project \cite{Br} to build  a 3 km detector in Italy, GEO-600 
project \cite{Ho} to build a 600 m detector in Germany and the
TAMA-300 project to build a 300 m  detector in Japan. The construction
of the above detectors has already  started.
First observations may come as early as 2000. These observations
have a potential to see the full range of gravitational wave sources: 
periodic, burst, quasiperiodic and stochastic.

In the development of data analysis techniques, it is useful to 
examine data already available from prototypes.
Today, prototype interferometers are routinely  operating at a 
displacement  noise level of a few times $10^{-19}$ m$/\sqrt{{\rm Hz}}$
over a frequency range from 200 Hz to 1000 Hz corresponding to an 
rms gravitational-wave amplitude noise level of 
$h_{rms}\sim 2\times 10^{-19}$ \cite{Abra,Ro}.

There are different noise sources  that limit the sensitivity of the
laser-interferometer detectors. 
The stochastic noise can be modeled as a sum of six main contributions
\cite{Kro}:
photon shot noise, seismic noise, quantum noise (which follows from
the indeterminacy of the position of the test masses due to the
Heisenberg uncertain principle),
the vibration of the suspension wires (\lq violin modes'),
and thermal noise from 
the vibration of the test masses and from the low frequency oscillations of the
pendulum suspensions.
The above noise sources may be considered among the most important  but
other sources  of noise cannot be ignored. 

In the measured noise spectrum of the different prototypes,
in addition to the stochastic noise, we observe
 peaks due external interference, where the amplitudes are not stochastic.
  The most numerous are  powerline frequency harmonics. We have shown 
  how to model and remove these very effectively using a technique we call 
  coherent line removal ({\sc clr}) \cite{AS1,AS2}. Other lines are clearly
  related to the power supply but appear at non-harmonic frequencies.
  In this paper, we
 will describe their characteristics and we will present a
procedure to remove them as well. Our goal is to remove as many interference
features as possible, so that the interferometer sensitivity 
is limited by the genuinely stochastic noise.

{\sc clr} is an algorithm  able to
remove interference present in the data while preserving the stochastic 
detector noise. {\sc clr} works when the  interference is present
in many harmonics and they remain coherent with each other. 
In \cite{AS1,AS2}, we applied {\sc clr} to some interferometric data
and the entire series of wide lines corresponding to the electricity
supply frequency and its harmonics were completely removed even when
the frequency of the supply was not independently known.

In addition to those lines appearing at multiples of the electricity supply
frequency (50 or 60 Hz), there are 
other interference lines whose frequencies change in step with the
supply frequency,
but not  at the harmonic frequencies. 
From a data-analysis point of view, we try to develop a technique able to
remove this interference while producing a  minimum disturbance to the 
underlying noise background. In order to remove  these lines,
other methods are also available \cite{Th,p3} but these methods will
remove the noise and any underlying real signals as well.

We are not sure if this type of interference will be absent in 
large-scale interferometers,
since the physical
process that generates all these lines is so far unknown.
But it is present in prototype data and therefore it is important to be
prepared to remove it from full data.

This paper is based on a study of  Glasgow interferometer data
taken in March 1996.
The method we propose  makes use of a reference wave-form signal
corresponding to the fundamental harmonic of the electrical interference.
We can  obtain it directly from the supply voltage or we can  construct it
using the {\sc clr} algorithm
 from the true harmonics present in the data.
The method we propose is an adaptive procedure that  is tuned in such a way 
that the electrical interference can be removed and \lq single-line' signals
masked by them can be recovered to at least  the $75\, \%$ level.

The rest of the paper is organized as follows: In section  II, we describe
the electrical interference present in the data.
In section III,  we present different models of the interference.
In section IV, we summarize the principle of the coherent line
removal algorithm and we explain how to construct a reference wave-form
of the incoming  electricity signal from the data.
In section V, we present an algorithm to remove  the 
electrical interference, and not only the harmonics of the reference
wave-form.  The algorithm is recursively
applied for small stretches  of data. Thus, it allows the parameters
to
change and to adapt themselves in order to be able to remove the
interference with a minimum  disturbance of the noise background.
Finally, in section VI, we discuss the results obtained.

\section{The electrical interference in the 
prototype data}

In this paper, we will focus our attention to the data 
produced by the Glasgow laser interferometer in March 1996.
The data set consists of 19857408 points, sampled at 4 kHz and
quantized with a 12 bit analog-to-digital  converter with a dynamical range
from -10 to 10 Volts. The data are divided into 4848 blocks of 4096 points
each. The first 18 minutes of data were rendered 
useless due to a failure of the autolocking. Thus, in our 
analysis  we ignore the first 1153 blocks. (For preliminary studies of
these data see \cite{J,K1,K,Soma}.)

In the study of the prototype data, we observe in the power spectrum
many instrumental lines. Some of them are  rather broad and
 appear at multiples of 50 Hz. But  there are other ones
appearing at different frequencies.
In the  data, the lines at 1 kHz have a width of 5 Hz. Therefore,
we
can ignore  these sections of the power spectrum or we can try to 
understand the interference  and, if possible, remove it
 in order to be able
to 
detect any possible gravitational wave signal previously masked by it.

We have already shown how to remove lines at integer multiples of the
supply frequency \cite{AS2}.
In long-term Fourier transforms, these lines are broad, and the structure
of different lines is similar apart from an overall scaling 
proportional to the frequency. In smaller length Fourier transforms,
the lines are narrow, with central frequencies that change with time,
again in proportion to one another. It thus appears
that all these lines are harmonics of a single source (e.g., the
electricity supply) and that their broad shape  is due to the wandering
of the incoming electricity frequency. These lines have been observed 
in different 
interferometer prototypes \cite{Abra,ga}.

But this is not the end of the story. Further analysis of the
prototype data reveals the presence of many other features that are
related to the incoming electricity frequency. These other lines
are not as powerful as the harmonics. In many cases, they are just
slightly above the stochastic noise level.

The easiest way to detect their presence is by studying in detail
the spectrogram (i.e., the magnitude of the time-dependent Fourier
transform versus time). This is a time-dependent frequency analysis
in which the whole data set is split into small segments and for each of 
them a discrete Fourier transform is computed. 
By examining the spectrogram, we identify a large
number of
lines that  present similar time-frequency evolution to the 
harmonics of 50 Hz. Therefore, all these line are related in some way
with the incoming electricity supply. An example is shown in figure 1.

These lines are spread over the whole spectrum. A big population of them
lies below 140 Hz. There are also some isolated ones around
222, 238, 444, 575, 887, 1105 and 1275 Hz and, after 1750 Hz, there is
again a large population of them. Our guess is that this kind of line
might be present through the whole spectrum, but many are buried in the
stochastic noise.

An interesting feature is that some of these lines appear to fall 
in to their own harmonic families.
 For example, we have found the   families:
(222, 443, 886, 1772 Hz), (48, 96 Hz), (66, 132 Hz), 
(72.5, 145 Hz), (116, 232 Hz). 
In all cases, the width of the lines (which is the interval of the frequency
wandering) seems to scale as the frequency. 
The amplitude of the lines  is time-dependent as well.

\begin{figure} 
\centerline{\vbox{ 
\psfig{figure=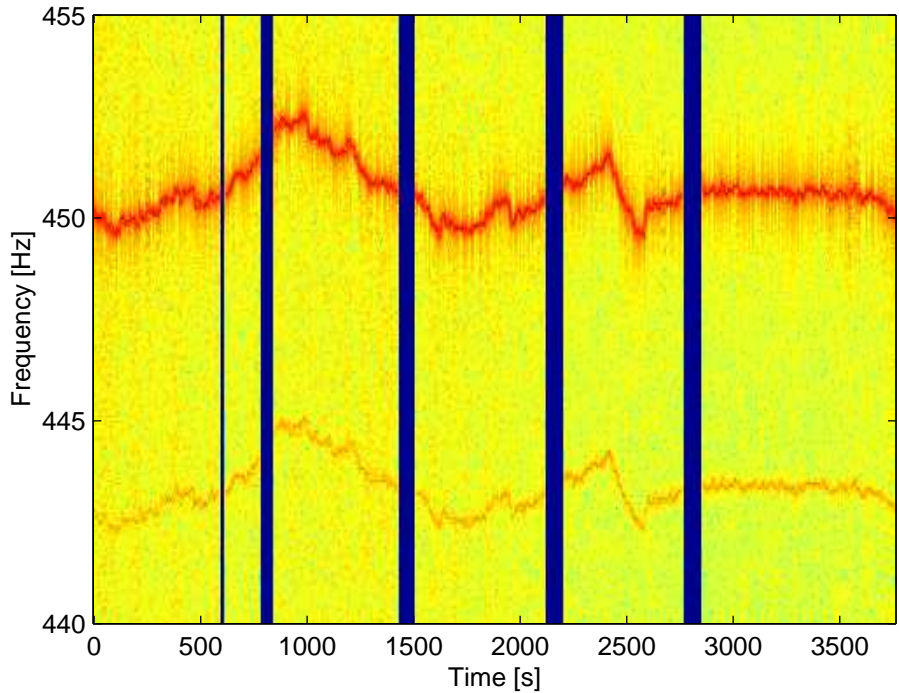,width=12.0cm} 
}} 
\vspace*{10pt}
\caption[]{Zoom of the spectrogram of the prototype data. The dark
areas correspond to the periods in which the detector is out of lock.
The line near 450 Hz  corresponds to the 9th harmonic of the
external electricity supply. The weaker line near 443 Hz is
one example of the many lines present in the spectrum that have a
similar time-frequency evolution to that of the harmonics of 50 Hz.
The frequency drift is
 due to the wandering of the incoming electricity 
frequency.
} 
\vspace*{10pt}
\end{figure} 

Although the nature of the process that generates these lines 
remains unknown, many of them may be intrinsic to the incoming
electricity signal. We have analyzed some a-posteriori 
electricity supply  voltage (from Glasgow University) and we 
have observed several features, and not just harmonics of 50 Hz.
External  switches or the running of other electrical devices
(as computers) can generate some of these lines. This could 
explain their non-linear and non-stationary character.
Another possible sources are ground motions due to 
mechanical motors running at different frequencies than 50 Hz
(e.g., trains), but whose frequency wanders in a way related 
somehow to the power supply.

Remembering the extremely small amplitude  of the disturbances 
these lines represent, it follows that it may be difficult for
experimenters to exclude these lines from the data of detectors
now under construction. It is therefore important to understand
how to remove them if they are there.

\section{Modeling the lines}

A first model for these non-harmonic lines is that they could be
beats between stationary frequencies and the supply harmonics.
We have examined this in detail and we believe it is unlikely. In
particular, the width of these lines seems to be proportional 
to their frequency, which could not be the case for a beat, and
also we would expect the lines to appear in couples, that it is
not the case.

Using this model,
\be
h(t)=\alpha M(t)^n \exp (i 2\pi f t ) \ ,
\ee
where $\alpha$ is a complex amplitude,
$M(t)^n$ is a supply harmonic,
and $f$ is the beat frequency,
 we have analyzed  several lines. For example,
the line at 99.7 Hz, we  find that
the best match corresponds to
 the second harmonic $n=2$ with a beat
frequency of $f=-0.7391$ Hz. Using these values,  we  try to 
remove the interference by applying a least square method, 
but as we show in figure 2 the interference is not cancelled.
The same occurs with other lines.
\begin{figure} 
\centerline{\vbox{ 
\psfig{figure=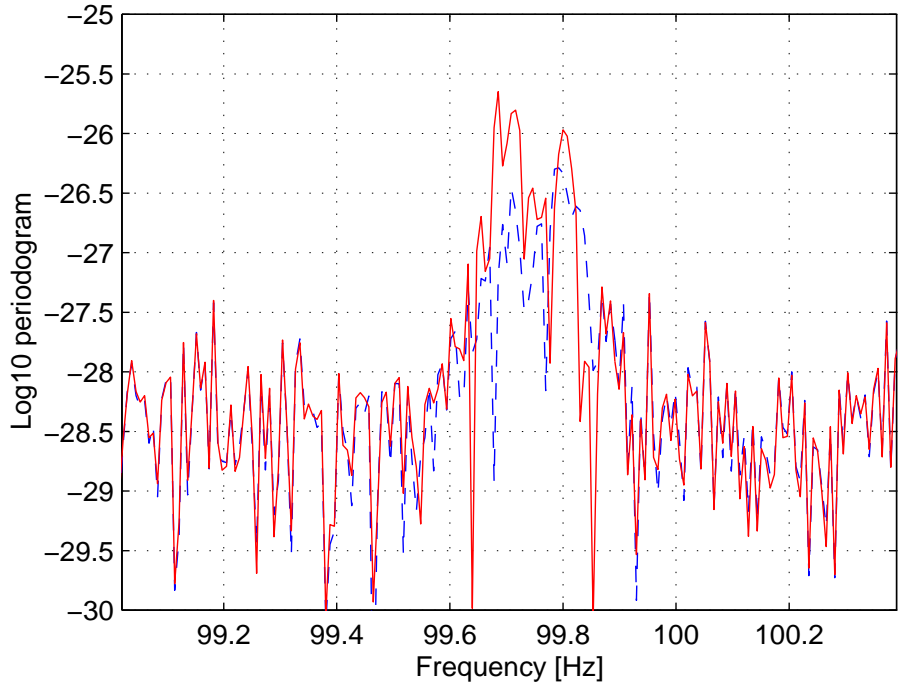,width=12.0cm} 
}} 
\vspace*{10pt}
\caption[]{Detail of the decimal logarithm of the periodogram.
 The solid line corresponds to 128 blocks
of the prototype data showing the line near 99.7 Hz.
The dashed line is the result after trying to remove that line
using the beat model.
} 
\vspace*{10pt}
\end{figure} 

After rejecting the possibility of the beats, the next
simplest model is  to assume a non-integer  harmonic of the supply:

\be
h(t)=\alpha (t) M(t)^q \ ,
\label{h1}
\ee
where $\alpha (t)$  is a slowly varying complex amplitude, 
$M(t)$ is a reference wave form 
corresponding to the fundamental harmonic of the electrical
interference, but now
$q$ is a real number, 
not only  an integer as in the case of the harmonics.
The reference signal, $M(t)$, can be obtained directly from 
the supply voltage, or it can be obtained directly using {\sc clr} as we
describe below. The justification for this model is simply the success
we have  in removing the interference.

The model is not perfect, and our  line-removal is sometimes incomplete.
It appears that the index $q$ may also be a slow function of time. 
But we will not  pursue this  refinement here.

\section{Coherent  Line Removal}
For any of the previous models,
we need to know the reference 
wave-form, $M(t)$, corresponding to the fundamental harmonic of the
interference. In the case of electrical interference, this wave-form can
be obtained  from the data applying the {\sc clr} algorithm.
In this section, we summarize the principle of {\sc clr}.
For further details we refer the reader to \cite{AS2}.

{\sc clr} 
 works when the interference is present in many harmonics,
and
assumes that the interference has the form
\be
y(t)=\sum_n a_n m(t)^n + \left( a_n m(t)^n\right)^* \ ,
\label{e3}
\ee
where $a_n$ are
complex amplitudes,  $m(t)$ is a nearly monochromatic function
 near
a frequency $f_0$ and $*$ is the complex conjugate.
The idea is to use
the information in the harmonics of the interference 
to construct the reference
 function $M(t)$ that is  as close a replica as possible to
$m(t)$.

Assuming additive noise, the data produced by the system
is just 
\be
x(t)=y(t)+n(t) \ ,
\label{e6}
\ee
where $y(t)$ is the interference
 given by Eq.~(\ref{e3}) and the noise $n(t)$ in the
detector  is a zero-mean stationary
stochastic process. The algorithm consists in defining
a set of functions $\tilde z_k(\nu)$
in the frequency domain as
\be
\tilde z_k(\nu)\equiv \left\{
\begin{array}{cc}
\tilde x(\nu) & \nu_{ik}<\nu <\nu_{fk}\\
0 & \mbox{elsewhere}\ ,
\end{array}
\right.
\label{e8}
\ee
where $\tilde{}$ denotes the Fourier transform,
 $(\nu_{ik}, \nu_{fk})$ correspond to the upper and lower frequency 
limits of the harmonics of the interference 
and $k$ denotes the harmonic considered.
 These functions are equivalent to
\be
\tilde z_k(\nu)= a_k \widetilde{m^k}(\nu) +\tilde n_k(\nu) \ ,
\ee
where $ \tilde n_k(\nu)$ is the noise in the frequency band of the
harmonic considered.  Their inverse  Fourier transforms yield
\be
z_k(t)=a_k m(t)^k +n_k(t) \ .
\ee
Since  $m(t)$ is supposed to be a narrow-band function near a frequency $f_0$,
each $z_k(t)$ is a  narrow-band function near $kf_0$. 
Then, we  define
\be
B_k(t)\equiv \left[ z_k(t)\right]^{1/k}\ ,\label{e10a}
\ee 
that can be rewritten as 
\be
B_k(t)= (a_k)^{1/k}m(t) \beta_k(t) \ ,
\ee
where
\be
\beta_k(t)=\left[ 1+ {n_k(t) \over a_k m(t)^k}\right]^{1/k} \ .
\label{e10}
\ee
All these  functions, $\{B_k(t)\}$, are almost monochromatic around the 
fundamental frequency, $f_0$, but they differ basically by a certain
complex amplitude. These factors, $\Gamma_k$, can easily be  calculated,
and  we can construct a set of functions   $\{b_k(t)\}$
\be
 b_k(t)=\Gamma_k B_k(t)\ ,
 \ee
such that, they all have the same mean value. Then,   $M(t)$ can be
 constructed  as a function of all $\{b_k(t)\}$
 in such a way
that it has the same mean and minimum variance. 
If 
$M(t)$ is linear with $\{b_k(t)\}$, then statistically the best choice for
$M(t)$ is
\be
 M(t)=\left(\sum_k {b_k(t) \over {\rm Var}[\beta_k(t)]} \right) {\Big
 { /}}
\left( \sum_k {1 \over {\rm Var}[\beta_k(t)]}\right) \ ,
\ee
where
\be
{\rm Var}[\beta_k(t)]= {\langle n_k(t) n_k(t)^*\rangle\over  k^2
\vert a_k m(t)^k\vert^2}+ \mbox{corrections} \ .
\ee
In practice, 
we  approximate
\be
\vert a_k m(t)^k\vert^2 \approx \vert z_k(t)\vert^2 \ ,
\ee
and we assume  stationary noise. Therefore, 
\be
\langle n_k(t) n_k(t)^*\rangle= \int_{\nu_{ik}}^{\nu_{fk}} S(\nu) d\nu \ ,
\ee
where $S(\nu)$ is the power spectral density of the noise.

The amplitude of the different
harmonics of the interference can be obtained then 
applying a least square method.

\section{Removal of non-harmonic external interference}

{\sc clr} can remove the integer 
harmonics of a reference signal but it does not
 remove other interference present in the data.
We can use  {\sc clr}  to construct a reference waveform 
$M(t)$, but we have to design another technique able to get
read off the electrical interference present at non-harmonics 
of the 50 Hz line frequency.

The  models 
 of  signals we propose are 
buried in noise. Therefore, we face  the problem of detecting signals and
estimating their parameters.

\subsection{Maximum likelihood detection}

A standard method is  the maximum likelihood 
detection which consists of maximizing the likelihood
function 
$\Lambda$ with respect to the parameters of the signal. If the maximum of
$\Lambda$ exceeds a certain threshold, we say that the signal is present.
(See \cite{Da,Sc1,Sc2} 
for signal analysis theory in the context of gravitational wave 
broadband detectors.)

We assume that the noise $n(t)$ in the detector is an additive,
zero-mean, Gaussian and stationary random process. Then the data
$x(t)$ (if the expected signal model  $h(t)$ is present) can be written as
\be
x(t)=n(t)+h(t) \ .
\ee
The logarithm of the likelihood function has the form
\be
\ln \Lambda=\sum_{k=0}^{N-1}{\tilde x_k \tilde h_k^*\over S_k} - {1\over 2}
\sum_{k=0}^{N-1}{ \vert \tilde h_k\vert^2\over S_k} \ ,
\label{h4}
\ee
where 
$S_k$ is the power spectral density of the noise, $k$ is the frequency index
running from $0$ to $N-1$ and $N$ is the number of sampled points.
The likelihood ratio $\Lambda$ depends on the particular set of data
$x(t)$ only through the sum
\be
G=\sum_k{\tilde x_k \tilde h_k^*\over S_k} \ .
\ee
This sum is called the detection statistic for the signal $h$. Its variance
is
\be
d^2=\sum_k{ \vert \tilde h_k\vert^2\over S_k} \ .
\ee
If there is  no signal present, the mean of $G$ is zero, but if the signal
$h$ is present the mean of $G$ will be equal to its variance.

The thresholds on the likelihood function $\Lambda$ must be set with regard
to the false alarm probability. For a detection statistics $G$ and a variance
$d^2$, the false alarm probability is
\be
P_F={1\over 2}{\rm erfc}\left( {G\over \sqrt{2} d}\right) \ .
\ee
This is equivalent to study the output signal-to-noise ratio ({\sc snr}),
i.e., the value of the detection statistics divided by the standard
deviation,
\be
{\sc snr}\equiv {G\over d}= \sum_k{\tilde x_k \tilde h_k^*\over S_k} {\Big /}
\sqrt{\sum_k{ \vert \tilde h_k\vert^2\over S_k}} \ .
\label{h5}
\ee
Notice that {\sc snr} (and therefore the false alarm probability) is
independent of the value of the amplitude of the model signal $h(t)$ used
for this pattern-matching procedure. Of course, the {\sc snr} is proportional
to the amplitude of whatever multiple of $h(t)$ is contained in $x(t)$ .
This is  thus a linear detector.

\subsection{The parameter space}

Assuming the  form for the electrical interference,
 $h(t)= M(t)^q$,
we have to construct as many filters as different values of $q$ need 
to be considered and, for each of them, calculate their {\sc snr}.

In order to set the
 parameter space, we can 
 consider $M(t)$ as a monochromatic signal at a frequency $f_0$.
Hence, $M(t)^q$ will be a monochromatic signal at $qf_0$.
Thus, the maximum value of $q$ to be considered corresponds to
\be
q_f={f_{Nyquist}\over f_0} \ ,
\ee
where 
\be
f_{Nyquist}={f_s\over 2} \ ,
\ee
 and $f_s$ is the sampling frequency.
 
The frequency resolution is $\Delta \nu=1/T$. Therefore,
 we can resolve two signals if the separation in $q$ is of the
order 
\be
\Delta q={\Delta \nu\over f_0} \ .
\ee
This is the maximum separation in $q$ we can allow.
 Note that the size of the parameter space 
\be
N_q={q_f\over \Delta q}={f_s\over 2} T \ ,
\ee
increases with the observation time $T$.

For 128 blocks of the Glasgow data (this corresponds to the usual stretch
 of data we work at once), we obtain
\be
\Delta q\leq 0.00015  \ .
\ee
We can  calculate the minimum number of filters and
the number of floating points operations  
 needed to calculate the {\sc snr}  for all of them. 

\subsection{An approximately matched filter}

The calculation of the exact matched filter for
all the parameter space of $q$ is  
computationally expensive.
Given the reference function $M(t)$, for each value of $q$,
we need to calculate $h(t)=M(t)^q$, perform its Fourier Transform, and
then, compute the output filter via the formula of the  {\sc snr}
given by Eq.~(\ref{h5}).

Since we assume $M(t)$ is a nearly monochromatic function,  all the
functions $h(t)=M(t)^q$ are also going to be nearly monochromatic but
at different frequencies. This means that  the values of their
Fourier transforms are just relevant in  small frequency bands. Hence, the 
first approximation we can perform is to reduce  the index summation in
Eq.~(\ref{h5})  to a small interval. This is equivalent to consider
 that the Fourier transform of the template is zero outside this interval.
 
 But this is not enough. What is really computationally expensive is the
 construction of the templates. Since the 
 reference signal $M(t)$ is almost monochromatic and   changes
 frequency smoothly in time, we can approximate the values of 
 $\widetilde{M^q}(\nu)$ in a certain interval, $\nu_{iq}<\nu<\nu_{fq}$,
 by those of $\widetilde{H^q}(\nu)$
 defined by
\be
\widetilde{H^q}(\nu)\equiv 
\widetilde{M^n}\left(\nu{q\over n}\right) \ ,
\label{h6}
\ee
where $n$ can be chosen to be an integer, i.e., we can build 
$\widetilde{H^q}$ as a similar function to 
$\widetilde{M^n}$, where $M^n(t)$ corresponds to an harmonic of the
reference signal.

If we construct $\widetilde{H^q}(\nu)$ via Eq.~(\ref{h6})
using the nearest harmonic, we will expect  
$\widetilde{H^q}(\nu)$ to be close to $\widetilde{M^q}(\nu)$.
Therefore, we just need to calculate the Fourier transforms of all the
harmonics (which is a small number in comparison to all the possible values 
of $q$), calculate $\widetilde{H^q}(\nu)$  from the
nearest harmonic and substitute their values in equation (\ref{h5}).

The power spectral density, $S_k$, can be estimated from the data using 
Welch Method \cite{w1} averaging over shorter periodograms.

With all these simplifications, we  apply the approximate
matched filter to 128 blocks (approximately two minutes) of prototype
data. By setting a threshold of {\sc snr}=4.5,  we have found
several values of $q$ for which the electrical interference
might be present. See table I.

\begin{center}
\begin{table}
\caption{Values of $q$ obtained from the approximate matched filter
with a {\sc snr} larger than 4.5, without including the harmonics.}
\begin{tabular}{cc}
$ q$ & {\sc snr}\\ \hline
0.66165  & 7.778\\
0.75515 & 4.609\\
 0.95895  & 6.317\\
   0.97865  & 5.295\\
   0.99505 & 8.875\\
   1.45185  & 5.614 \\
   1.91845  & 8.298 \\
   1.98480  & 12.849 \\
   1.99975  & 11.520 \\
   2.26690  & 5.015 \\
   2.31580  & 5.197 \\
   2.49535  & 7.656 \\
   2.55305  & 4.626 \\
   2.64640  & 4.875\\
   4.42645  & 7.093 \\
   4.75555  & 12.981 \\
   8.85255  & 7.665 \\
  17.70450  & 4.516 \\
  26.96785  & 6.497\\
  35.75285  & 6.539\\
  36.96795  & 8.202
\\ 
\end{tabular}
\end{table}
\end{center}

If we assume that the 
interference follows the  model
\be
h(t)=\alpha M(t)^q \ ,
\ee
with $\alpha $ being constant,
for each $q$ we can find the value of the complex amplitude
$\alpha$ by applying a least square method in the time domain, i.e.,
$\alpha$ is the value of $\beta$ that minimizes 
 the quantity  $\vert x(t) -\beta M(t)^q\vert^2$.

With this method and
 using the same 128 blocks of data, we
try to remove the interference. 
The result
is that the interference is 
attenuated but not removed. 
Therefore, we think that 
this model is too simple
and that to  remove  real 
interference we need a more complicated model.

\subsection{Amplitude modulation}

The extra complication is to allow the amplitude to vary slowly with time.
Therefore, we assume that the interference takes the form
\be
h(t)=\alpha (t)M(t)^q \ ,
\ee
where $q$ are real constants, $\alpha (t)$ is a slow changing
function of time, and the values of $q$ do not differ  much
from those calculated with the matched filter as described 
in the previous subsection.

We have to find a 
procedure to determine $\alpha (t)$, the
 amplitude modulation. Since the signal is buried
in noise, we cannot find its exact value. What we do is to find
an approximate  function $\beta(t)$ to  $\alpha (t)$ such that allows
us to remove the interference and, at the same time, keep the intrinsic level
of noise present in the interferometer.

The method we use consists of splitting the data into small pieces 
(i.e., small number of blocks), and for each piece calculating a value
$\beta$  as if it was a constant. Then, we construct $\beta(t)$ as the
succession of those values obtained.

In practice, we separate the data into sets of $2n_b+1$ blocks with
overlaps of $n_b$ blocks. For each of them  we calculate  the $\beta$
value and we associate  it to the 
block $n_b+1$. In this way, we construct $\beta(t)$
as a set of discrete values that  change smoothly in time.
The number of blocks, $n_b$, used must be tuned according to the data.
If the value of  $n_b$ is too big, it does not allow enough amplitude
modulation and the interference is not cancelled. By contrast, if
 $n_b$ is too small, the function $M(t)^q$ for those small number of
 blocks will be almost monochromatic, in the sense that it will
 affect to a few frequency bins. Then, this method  will behave
 as an adaptive  multitaper method \cite{Th,p3} and hence, it will 
 remove whatever is in those frequency bins (any signal plus noise).

For the prototype data, we have found that these values, $n_b$, need to depend
on $q$. Thus, we build the  function $n_b(q)$ satisfying the following
 requirements:
\begin{itemize}
\item it must be able to remove the interference,
\item it should  leave a noise level comparable to the intrinsic 
noise background of
the interferometer,
\item  an artificial \lq single-line' signal  with an
amplitude of the order of the electrical interference
should not be attenuated more than  25 \%.
\end{itemize}
To satisfy all these requirements, we propose the following function
\be
n_b(q)=\left\{\begin{array}{l c}
7 \ , & q<2\\
5\ , & 2\leq q<4\\
4\ , & 4\leq q<6\\
3\ , & 6\leq q<8\\
2\ , & 8\leq q<16\\
1 \ , & q\geq 16 \ .
\end{array}
\right.
\ee
Using  this function, $n_b$, and the values of $q$ displayed in table I,
we have  succeeded in removing
 the interference present  in 128 blocks. The result are shown in figures 3
 and 4.

\begin{figure} 
\centerline{\vbox{ 
\psfig{figure=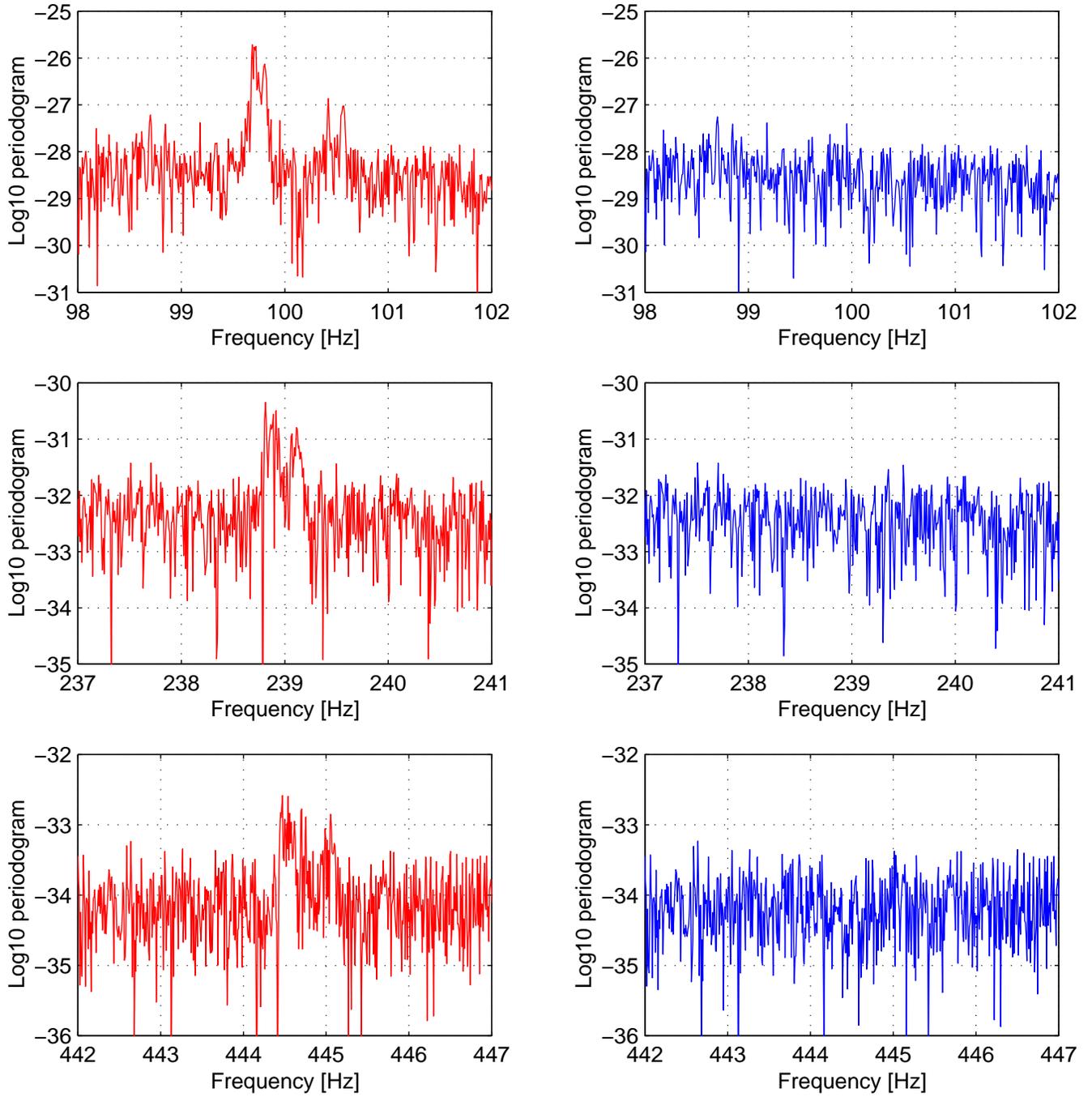,width=7.0in} 
}} 
\vspace*{10pt}
\caption[]{Decimal logarithm of the periodogram 
of $2^{19}$ points (128 blocks) of the prototype data. 
 (Left) Details of the lines
near 99.5, 100.5, 239 and 445 Hz. (Right) The same data after removing
the electrical interference as described in the text.
} 
\vspace*{10pt}
\end{figure} 
\begin{figure} 
\centerline{\vbox{ 
\psfig{figure=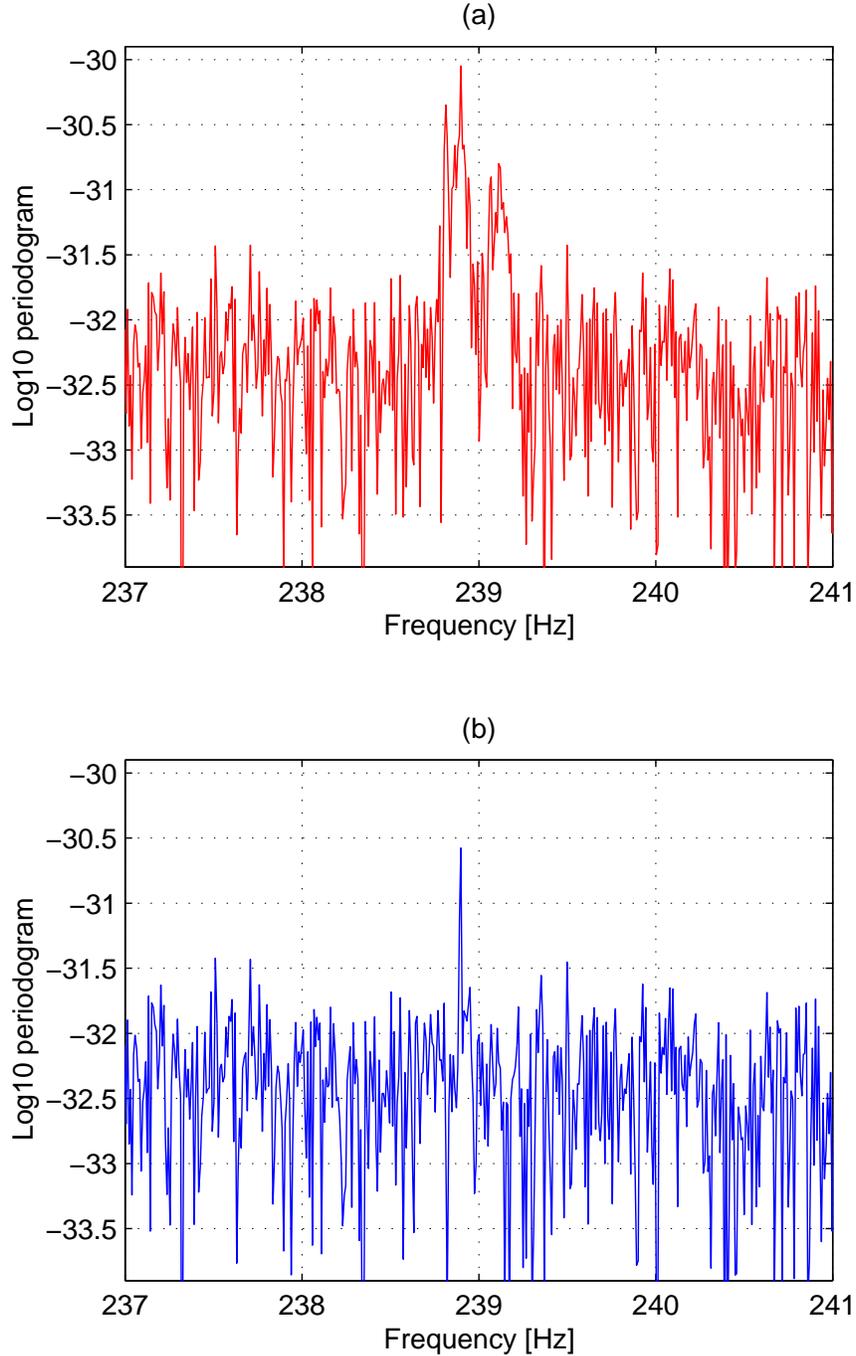,width=4.5in} 
}} 
\vspace*{10pt}
\caption[]{
 (a) The same experimental data as in figure 3 with an 
  artificial signal added at 238.9 Hz.
  (b) The  data in (a) after removing
the  interference, revealing that single line signals can be recovered.
} 
\vspace*{10pt}
\end{figure} 

\subsection{Frequency drift}

 We proceed now to remove the interference  from the whole data
 stream.
 We split the data into fragments of 128 blocks
 and we apply the previous method using the amplitude modulation.
 
As a first attempt, we  assume that the values of $q$
remain constant during all time. As a result, some of the lines
are removed but some others are not after a certain time.
 
This is not surprising. Visual  inspection of the spectra shows that
these lines drift,
\be
q(t)=q_0+\delta q(t) \ ,
\ee
where $\delta q(t)$ is  small in comparison with $q_0$.

Therefore, for the longer data set, 
 we  allow small changes in $q$.
 We apply to each fragment of 128 blocks the 
  approximate matched
filter, as described before, but using a much reduced parameter space,
i.e., just  allowing
a maximum variation of $10\ \Delta q$ about the $q$ values of the 
previous fragment of 128 blocks.

In order to construct the templates, we make use again of Eq.~(\ref{h6}),
but we use the old values of $q$ (those obtained
in the previous data fragment).
Then we  choose the new values of $q$ by  maximizing the
{\sc snr}.

Using this procedure, we have  removed all the lines
corresponding to those initial values of $q$ listed in  table I.
See figure 5.
\begin{figure} 
\centerline{\vbox{ 
\psfig{figure=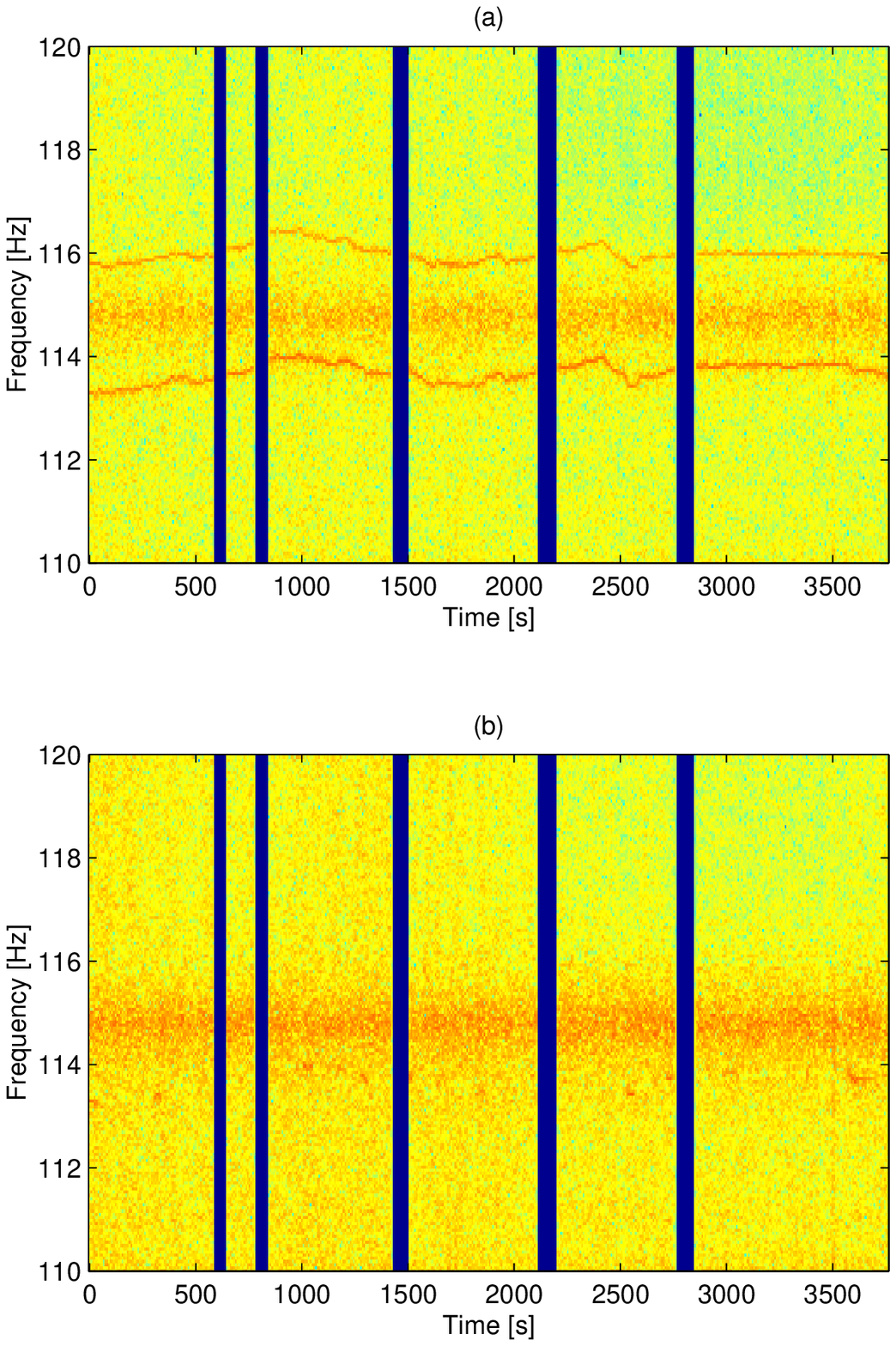,width=12.0cm} 
}} 
\vspace*{10pt}
\caption[]{Comparison of a zoom of the spectrogram.
(a) is obtained from the prototype data. We see two anomalous
lines near 114 and 116 Hz. (b) The same spectrogram as in (a)
after removing the interference using variable values of $q$.
} 
\vspace*{10pt}
\end{figure} 

From the evolution of $q$, we observe that some values of $q$ remain
almost constant, but many others change in time, obtaining a 
maximum variation of 0.005 for 3695 blocks of data.

Assuming the incoming signal is monochromatic, we can estimate
the best timescale to perform the matched filter, i.e., the length
of the  stretch of data  for which the frequency resolution is greater than
the maximum expected frequency variation due to the drift of $q$.
This yields 120 blocks, while we were using 128.

 \section{Discussion}
 
We have described an algorithm able to remove any kind of interference
related with the incoming  electricity main supply.
The study is based on the data produced by the Glasgow interferometer
prototype. In the data, we have observed many interference lines
that are highly non-linear and non-stationary.

The form of the interference can be modeled by $h(t)=
\alpha (t)M(t)^{q(t)}$, where $M(t)$ corresponds to the 
fundamental harmonic  of the incoming electrical signal,
$\alpha (t)$ is a slow varying function of time, and 
$q(t)$ is almost constant, but it can drift in time, i.e.,
$q(t)=q_0+\delta q(t)$, where $\delta q(t)$ is small in
comparison to $q_0$.

In order to detect those signals and estimate their parameters,
we use  the method of  maximum likelihood detection to
determine the values of $q$ (at a certain instance). Then,
we apply an adaptive procedure to determine the amplitude modulation and
we repeat it recursively for the whole data stream.

The result is that all those lines which were initially detected 
(i.e., those with enough {\sc snr}) have been  tracked and 
completely removed.
This method is able to recover monochromatic
signals that are buried by the interference. The  signal distortion
is  less than  $25\, \%$. Thus,
this procedure can assist in the search for 
continuous waves and also clean the statistics of the noise in the
time domain.
As we pointed out  in an earlier paper \cite{AS2}, the removal of 
 lines like these can reduce  the level of non-Gaussian noise. 
Therefore, line removal is important  since it can raise the 
sensitivity and duty cycle of the detectors  to short bursts of
gravitational waves, as well.
 
\acknowledgements{
We would like to thank 
J. Hough and the gravitational waves group at Glasgow University
for providing their gravitational wave interferometer data for analysis.
This work was partially supported by the European Union, 
 TMR Contract
No. ERBFMBICT972771.}


\end{document}